\documentclass[11pt,tightenlines,aps,prd,groupedaddress]{revtex4}

\usepackage{amsmath,amssymb,bm}

\usepackage{graphicx}
\usepackage{amsfonts}
\usepackage{amssymb}
\usepackage{amsbsy}
\usepackage{amsmath}
\usepackage{latexsym}
\usepackage{pdfpages}
\usepackage{bm}
\usepackage{color}
\usepackage{comment}
\usepackage[colorlinks=true,breaklinks]{hyperref}

\newcommand*{\di}{\partial}

\newcommand{\eq}[1]{(\ref{#1})}
\newcommand*\bigcdot{\mathpalette\bigcdot@{.5}}

\def\l{\left}
\def\r{\right}
\def\p {\partial}

\def\be {\begin{eqnarray}}
\def\ee {\end{eqnarray}}
\def\nn {\nonumber}

\def\eps{\epsilon}
\def\abar{\bar{a}}
\def\pbar{\bar{p}}

\def\tphi{\tilde{\phi}}
\def\tp{\tilde{p}}

\begin{document}

\title{Cosmological perturbation theory with matter time}

\author{Viqar Husain}
\author{Mustafa Saeed}

\email{vhusain@unb.ca}
\email{msaeed@unb.ca}
\affiliation{Department of Mathematics and Statistics\\
University of New Brunswick\\
Fredericton, NB E3B 5A3, Canada}


\begin{abstract}
 
We study cosmological perturbation theory with scalar field and pressureless dust in the Hamiltonian formulation, with the dust field chosen as a matter-time gauge.   The corresponding canonical action describes the dynamics of the scalar field and metric degrees of freedom with a non-vanishing physical Hamiltonian and spatial diffeomorphism constraint. We construct a momentum space Hamiltonian that describes linear perturbations, and show that the constraints to this order form a first class system. We then write the Hamiltonian  as a function of certain gauge invariant canonical variables and show that it takes the form of an oscillator with time dependent mass and frequency  coupled to an ultralocal  field.  We compare our analysis with other Hamiltonian approaches to cosmological perturbation theory that do not use dust time. 

\end{abstract}

\maketitle

\section{Introduction}

The diffeomorphism symmetry of  general relativity (GR) is manifested in its canonical formulation through the presence of phase space constraints that generate a closed poisson algebra. The hamiltonian is a linear combination of these constraints, and so vanishes on shell \cite{Arnowitt:1962hi, Hanson:1976cn}. The path to a physical non-vanishing hamiltonian requires selecting a function on the phase space as a choice of time;  the negative of the phase space variable conjugate to the time choice provides  this hamiltonian\cite{KUCHA__2011}. It is clear that there are numerous choices for physical hamiltonians, and the classical dynamics generated using these, for given ansatze, leads ultimately to the same solutions, but in different charts, and covering different regions  of the spacetime manifold.

In early work on general  relativity, time choices were divided into ``intrinsic," where time is a function of the spatial metric, and ``extrinsic," where time is a function of the extrinsic curvature (or the momentum conjugate to the spatial metric). Two frequently studied examples of such choices are 3-volume, and trace of the extrinsic curvature (York time) \cite{York:1972sj}. 

For GR with matter fields, there is also the possibility of using matter phase space functions as clocks. Examples of such clocks date back to early studies of cosmological models where a scalar field was used as a clock \cite{Blyth:1975is}.  More generally Brown and Kuchar   \cite{Brown:1994py} gave a prescription for using a 4-component  fluid field coupled to GR as a matter reference system for time and space. This idea, along with the older one of scalar field time, has subsequently been used in many works with the aim of building models for quantum gravity \cite{Husain:2011tk,Husain:2011tm,Giesel_2015,Assanioussi_2017}.

There are two closely related approaches in which geometric or matter reference systems may be used for classical and quantum models. One of these is to fix the gauge and solve the corresponding constraint strongly and thereby obtain a partially or fully gauge fixed (or ``deparametrized") system. The other is to use ``relational" observables \cite{Tambornino:2011vg} without deparametrizing, where the evolution of one variable is observed relative to that of the chosen clock variable. This latter procedure generates gauge invariant (Dirac) observables through eliminating the arbitrary time parameter $t$ by inverting the evolution of a clock phase space variable  T:  one inverts $T(t)\rightarrow t(T)$ in some domain, and then  substitutes $t(T)$ into any other observables ${\cal O}(t)$ of interest, $ {\cal O}(t) \rightarrow {\cal O}(t(T))$. 

In this paper we apply a specific matter time gauge -- dust time --- to cosmological perturbation theory in the Hamiltonian formulation. The principal advantage of this gauge is that the physical hamiltonian takes a simple form: it is exactly the same algebraic expression as the hamiltonian constraint. At this first stage we do not fix the spatial diffeomorphism symmetry, which remains as a decoupled gauge symmetry until we proceed to a second order expansion of the canonical action.  

Our work is not the first to construct a hamiltonian perturbation theory for cosmology. The first such analysis was given in \cite{Langlois:1993}; others using the relational approach have appeared recently\cite{Giesel_2019,Giesel:2020bht}. However our approach differs from both in several respects, the primary one being that we use only a  clock field, and fix a matter-time gauge strongly at the outset before proceeding to cosmology \cite{Husain:2011tk}. This step simplifies the analysis significantly by removing the hamiltonian constraint at the outset.  Another important difference is that the theory we consider, GR, with dust and scalar field, has four local physical degrees of freedom, two gravitational, one scalar and one dust. Therefore  after selecting the dust time gauge, the metric acquires an additional degree of freedom.  In these aspects our work complements these earlier works, with little overlap. The approach we follow was used by one of the authors for studying perturbations on Minkowski space \cite{Ali_2016}; the work presented here may be considered an extension of this to cosmological perturbation theory. It may be generalized to include additional matter fields.
  
 This work may also be viewed in a wider context of GR coupled to special types of matter. These include  the Einstein-Aether models \cite{Eling:2004dk}, where a dynamical vector field of timelike norm is added to the GR action. A linearized analysis of these models has been performed, with the result that the graviton modes decouple from the aether modes \cite{Jacobson:2004ts}.  The other model is the  so-called mimetic gravity \cite{Chamseddine:2013kea, Golovnev_2014}, where the conformal mode of the spacetime metric is encoded as a scalar field with an arbitrary potential. This extra mode in the gravitational field represents  self-interacting matter with arbitrary  potential \cite{Lim:2010yk, Chamseddine:2014}, and has been used to model inflationary and bouncing cosmologies \cite{Chamseddine_2017}.  Given these analogies, it is  potentially useful to consider this  work in the larger context of Einstein-Aether \cite{Jacobson:2015mra} and mimetic gravity theories. Indeed, the dust-time gauge we employ here may be considered a natural choice for all scalar-tensor theories of gravity, among which Einstein-Aether and mimetic gravity are but two examples. 

In the next section we review the use of the dust time gauge in the ADM canonical framework \cite{Brown:1994py, Husain:2011tk}.  In Sec. III we develop the linearized perturbation theory by expanding the hamiltonian and diffeormorphism constraints about an arbitrary FLRW-scalar solution. We show from the canonical perspective that the graviton equations turn out to be exactly those derived  in the standard covariant perturbation theory without dust (see e.g. \cite{Baumann:2012}), and that the vector modes may be gauged away. In Sec. IV we introduce diffeormorphism invariant phase space variables to study the scalar field and curvature degrees of freedom (which are independent degrees of freedom in the dust time  gauge). In Sec. V we  give a detailed comparison with standard perturbation theory. We conclude in Sec. VI with a summary and possible future directions. Several appendices provide details of our calculations: Appendix \ref{HA} gives details of the Hamiltonian perturbation expansion, Appendix \ref{gravitonA} gives a derivation of the graviton equation, and Appendix \ref{diffA} provides a proof that the linearized constraints are first class.
 
\section{Hamiltonian gravity with dust}
 
We consider GR coupled to dust and a scalar field. The action is 
\be
S= -\frac{1}{2\pi}\int{d^{4}x\sqrt{-g}R} +\frac{1}{4\pi}\int{d^{4}x\sqrt{-g}\ m(g^{ab}\di_{a}T\di_{b}T+1)}  + \int d^4x\  {\cal L}(\Phi).
\ee
The  second term is the dust action, and the last term  is the minimally coupled scalar field with an arbitrary potential  $V(\Phi)$.  With $u^a = \di_a T$, the dust energy-momentum  tensor is
\be
  T_D^{ab} = m u^au^b + \frac{m}{2} g^{ab} \left(g_{cd}u^cu^d +1 \right).
\ee
Thus on shell,  $m$ is interpreted as the dust energy density.

The ADM canonical theory obtained from this action is 
\be
S= \int{dt \ d^{3}x\left(\pi^{ab}\dot{q}_{ab}+p_{\Phi}\dot{\Phi} + p_T \dot{T}-N\mathcal{H}-N^{a}\mathcal{C}_{a}\right)},
\label{can-act}
\ee
where the pairs  $(q_{ab},\pi^{ab})$, $(\Phi, p_{\Phi})$ and $(T, p_T)$ are respectively the phase space variables of gravity, scalar field and dust.  The lapse and shift functions,  $N$ and $N^{a}$ are the coefficients of the Hamiltonian and diffeomorphism constraints
\be
\label{HG}
\mathcal{H} &=&\mathcal{H}^{G}+\mathcal{H}^{D}  + \mathcal{H}^{\Phi},\\
\mathcal{C}_{a}&=&\mathcal{C}^{G}_{a}+\mathcal{C}^{D}_{a} +\mathcal{C}^{\Phi}_{a}\nn\\
 &=&-2D_{b}\pi^{b}_{a}+ p_T\di_aT + p_{\Phi}\di_{a}\Phi  ,
\ee
where  
\be
\mathcal{H}^{G} &=& \frac{1}{\sqrt{q}} \left( \pi^{ab}\pi_{ab} - \frac{1}{2} \pi^2  \right) - \sqrt{q}R^{(3)} \\
\mathcal{H}^{D} &=&\frac{1}{2} \left[ \frac{p_T^2}{m\sqrt{q}}+m\sqrt{q} \left(q^{ab}\di_{a}T\di_{b}T+1\right)\right]\\
\mathcal{H}^{\Phi} &=& \frac{1}{2}\left(\frac{p_{\Phi}^2}{\sqrt{q}} + \sqrt{q} q^{ab}\di_a\Phi\di_b\Phi \right),
\ee
$\di_a,D_a$ are the spatial partial and covariant derivatives, and $R^{(3)}$ is the spatial Ricci scalar.  
The field $m$ appears only in ${\cal H}^D$  as an auxiliary field. We can therefore solve its e.o.m. for $m$ and substitute the result back into 
$\mathcal{H}^{D}$: 
\be
\label{m}
m=\pm\frac{p_T}{\sqrt{q(q^{ab}\di_{a}T\di_{b}T+1)}}.
\ee
\be
\mathcal{H}^{D}=  \text{sgn}(m)\  p_T  \sqrt{q^{ab}\di_{a}T\di_{b}T+1}. \label{Hd}
\ee
With this expression for $\mathcal{H}^{D}$, the final canonical action retains the form  \eq{can-act}, but now with no dependence on $m$ except for its sign.


\subsection{Dust time gauge}

We now introduce a partial gauge fixing by setting a time gauge to obtain a physical Hamiltonian; this fixes the time-reparmetrization invariance, while the spatial diffeomorphisms  remain as a full gauge symmetry. We use the dust time gauge  \cite{Husain:2011tk,Swiezewski:2013} which equates the physical time with the dust field, i.e., the spatial hypersurfaces are level surfaces of the dust field,
\be
\label{gauge}
\lambda\equiv T- \epsilon t \approx 0, \ \ \  \epsilon=\pm 1.
\ee
This is a special case of the Brown-Kuchar matter reference frame system which is designed to fix all four coordinate conditions.
The condition (\ref{gauge}) has a nonzero Poisson bracket with the Hamiltonian constraint, so this pair of conditions  constitute a second class set. According to the Dirac criteria,  a gauge condition is considered suitable if the matrix of second class constraints is invertible at all points \cite{Hanson:1976cn}.  In the present case  this matrix is   
\be
\label{matrix}
C = \l[ \begin{matrix}  0 & \{\lambda, \mathcal{H}\} \\  \{\mathcal{H}, \lambda \} &0 \end{matrix} \r] =  \text{sgn}(m) \l[ \begin{matrix} 0 & 1 \\ -1 & 0 \end{matrix} \r].
\ee
This matrix is invertible everywhere on the manifold. Therefore the dust time gauge does not breakdown at any point and is therefore a robust choice. The second condition on a canonical gauge is that it be preserved  in time. This gives an equation for the lapse function:
\be
\epsilon = \dot{T}= \left. \left\{T, \int d^3x \left (N  \mathcal{H}  + N^a \mathcal{C}_{a}\right) \right\} \right|_{T=t}  =   \text{sgn}(m) N\ . \label{gauge-pres}
\ee

Solving the Hamiltonian constraint for $p_T$ and substituting the gauge condition back into \eq{can-act} gives the gauge fixed action
\be
S_{GF} =  \int{dt \ d^{3}x\left[ \pi^{ab}\dot{q}_{ab}+p_{\Phi}\dot{\Phi} - \epsilon\ \text{sgn}(m)\left( \mathcal{H}^{G} +\mathcal{H}^{\Phi}\right) - N^{a}\mathcal{C}_{a}\right]}, \label{GF-act}
\ee
This identifies the physical Hamiltonian density 
\be
{\cal H}_P = \epsilon\ \text{sgn}(m)\left( \mathcal{H}^{G} +\mathcal{H}^{\Phi} \right) =   \text{sgn}(N)\left( \mathcal{H}^{G} +\mathcal{H}^{\Phi} \right), \label{hp}
\ee
where the last equality follows from \eq{gauge-pres}. Thus the physical Hamiltonian is determined up to an overall sign of the lapse function. Since we are free to choose the lapse up to sign, we will work with $N=1$. The corresponding spacetime metric is 
\be
ds^2 = -dt^2 + (dx^a + N^a dt)(dx^b + N^b dt) q_{ab}. 
\ee 

In the following we apply the dust-time canonical action \eq{GF-act} to flat FLRW cosmology and construct the linearized perturbation theory. At this stage we note the  central difference with standard perturbation theory: we have a physical Hamiltonian not a Hamiltonian constraint, therefore the gauge invariant  observables  we work with are those that are invariant under the spatial diffeomorphisms. Furthermore, the physical Hamiltonian \eq{hp} is what would be the Hamiltonian constraint for the gravity-scalar system. As a result, per point we have three physical degrees of freedom in the metric, and one in the scalar field;  the presence of the third degree of freedom in the metric is due ultimately to the fact that our starting action had a dust field.   As we will show, these can be rearranged into two graviton modes, a curvature perturbation, and the scalar field, with a relatively simple coupled dynamics. 

 \section{Cosmological perturbation theory}
 
 Our starting point for developing a canonical perturbation theory for flat FLRW models is the selection of a background solution starting with the action \eq{GF-act}. This starting point is distinct from all standard treatments of the subject, both canonical and covariant, with  the key difference being that the Hamiltonian constraint is no longer a  constraint, but is instead the physical Hamiltonian. This has several consequences, the main one being that the additional local degree of freedom that came from the dust  emerges in the metric perturbation.  
 
 Let us take the following parametrization for the ADM variables for the background solution:
 \be
q_{ab}^{(0)}&=&\bar{a}^{2}(t)\ e_{ab}, \quad    \pi^{ab (0)} =\left(\frac{\bar{p}(t)}{6\bar{a}(t)} \right)  e^{ab}\\
\Phi^{(0)} &=& \bar{\phi}(t), \quad \quad \quad p_{\Phi}^{(0)}= \bar{p}_\phi(t)\\
N^{a (0)}&=&0 
 \ee
where $e_{ab}$ is the Euclidean metric, $(\bar{a}(t),\bar{p}(t))$ and $(\bar{\phi}(t), \bar{p}_\phi(t))$ are the scale factor and scalar field and their conjugate momenta.  Substituting these into the dust-time gauge fixed canonical action \eq{GF-act} gives  the reduced action for the background 
\begin{equation}
S=\int{dt\left[\dot{\bar{a}}\bar{p}+\dot{\bar{\phi}} \bar{p}_\phi- \bar{\cal{H}}\right]}
\end{equation}
where 
\begin{equation}
\bar{\cal{H}}=-\frac{\bar{p}^2}{24\bar{a}}+\frac{\bar{p}_{\phi}^{2}}{2\bar{a}^{3}}+\bar{a}^3 V(\bar{\phi}).  
\end{equation}
 The background spacetime metric with this parametrization, with $N^2=1$ in the dust time gauge, is of the standard form
\be
ds^2 = -dt^2 + \bar{a}^2(t)e_{ab}dx^adx^b.
\ee 
The background equations of motions are 
\begin{subequations}
\label{eq:zeroeqns}
\begin{align}
\dot{\bar{a}}&=-\frac{\bar{p}}{12\bar{a}}\\
\dot{\bar{p}}&=-\frac{\bar{p}^{2}}{24\bar{a}^{2}}+\frac{3\bar{p}_{\phi}^{2}}{2\bar{a}^{4}}-3\bar{a}^{2}  V(\bar{\phi})  \\
\dot{\bar{\phi}}&=\frac{\bar{p}_{\phi}}{\bar{a}^{3}}\\
\dot{\bar{p}}_{\phi}&= -\bar{a}^{3}V'(\bar{\phi}),
\end{align}
\end{subequations}
where $V'(\bar{\phi}) = dV/d\phi |_{\bar{\phi}} $. The physical Hamiltonian is a constant of the motion in the dust time gauge, since it is not explicitly time dependent; this is unlike other time gauges, such as volume time $a^3=t$.  The background solutions therefore fall into 3 classes, $\bar{\cal H}=0$,   $\bar{\cal H}>0$, and $\bar{\cal H}<0$. The first of these corresponds to  the condition
\be
\frac{\bar{p}^2}{24\bar{a}}=\frac{\bar{p}_{\phi}^{2}}{2\bar{a}^{3}}+\bar{a}^3 V(\bar{\phi}),
\ee 
which by the equation of motion for $\abar$ (and restoring the $8\pi G$ factor) is the Friedmann equation
\be
\bar{H}^2 = \frac{8\pi G}{3}  \left( \frac{\bar{p}_{\phi}^{2}}{2\abar^6}+ V(\bar{\phi}) \right),
\ee
 where $\bar{H} = \dot{\abar}/\abar$. For the cases $\bar{\cal H} = \mu = $ constant $\ne 0$, the conservation of the physical Hamiltonian may be written 
 \be
 \bar{H}^2 = \frac{8\pi G}{3}  \left( \frac{\bar{p}_{\phi}^{2}}{2\abar^6}+ V(\bar{\phi}) -\frac{\mu}{\abar^3}\right), 
 \ee
 which shows that $\mu$ gives the dust energy density contribution to the Friedmann equation. This completes our summary of the background solutions in the dust time gauge. 
\subsection{Linearized Theory}
  
 We define the following expansion of phase space variables and the shift vector:
\begin{subequations}
\label{eq:genans}
\begin{align}
q_{ab}(t,\vec{x}) &= \bar{a}(t)^{2}e_{ab} + h_{ab}(t,\vec{x})\\
\pi^{ab}(t,\vec{x}) &= \frac{\bar{p}(t)}{6\bar{a}(t)}e^{ab} +  p^{ab}(t,\vec{x})\\
N^{a}(t,\vec{x}) &=  0+ \xi^{a}(t,\vec{x})\\
\Phi(t,\vec{x}) &= \bar{\phi}(t)+ \phi(t,\vec{x})\\
p_{\Phi}(t,\vec{x})&=\bar{p}_{\phi}(t)+p_\phi(t,\vec{x}).
\end{align}
\end{subequations}
Here the fields $h_{ab}, p^{ab}, \phi, p_\phi$ are respectively the perturbations of the gravitational and scalar field phase space variables, and $\xi^a$ is the perturbation of the shift vector.  
 
These are substituted into the physical hamiltonian and  spatial diffeomorphism constraint, which are then expanded to second order in the perturbations. This leads to  the second order action for the perturbations
\be
S^{(2)}=\int{dt d^{3}x\left[\dot{h}_{ab}p^{ab}+\dot{\phi}p_\phi - {\cal H}^{(2)} - \xi^{a}C_{a}^{(1)}\right]}, \label{pert-action}
\ee
 where ${\cal H}^{(2)}$  is the second order  perturbation of the Hamiltonian, and  $C_{a}^{(1)}$ is first order perturbation of the spatial diffeomorphism constraint. The latter is all that is required since the shift is first order. We note also that terms linear in the perturbations vanish when the background solution is imposed; the first order symplectic term in the action combines with the first order term ${\cal H}^{(1)}|_{\bar{S}}$ to give zero,  and the first order diffeomorphism term $\left(N^{a(0)} C_{a}^{(1)} + \xi^a C_{a}^{(0)}\right)|_{\bar{S}} =0$. ($\bar{S}$ denotes evaluation on the background solution.)  The expressions for 
 \be
\label{hgr2}
 {\cal H}^{(2)} &=& \mathcal{H}^{G(2)}+\mathcal{H}^{\Phi(2)}\\
  C_{a}^{(1)} &=&  C_{a}^{G(1)} + C_{a}^{\Phi(1)}
  \ee 
  are the following: 
 \be 
\label{eq:hgr2}
\mathcal{H}^{G(2)}&=& \abar\left( p^{ab}p_{ab} -\frac{1}{2} p^2 \right) + \frac{1}{\abar}\left(\frac{\pbar}{6\abar}\right) \left( p^{ab}h_{ab} - \frac{1}{2}hp \right) + \frac{1}{8\abar^3} \left( \frac{\pbar}{6\abar}\right)^2 \left( 5h_{ab}h^{ab} -\frac{3}{2} h^2  \right) \nn \\
&&- \frac{h}{2\abar^{3}}\left(\p_a\p_b h^{ab} - \frac{1}{2} \p^2h  \right) 
+ \frac{h^{ab}}{2\abar^{3}} \left(\p_b\p^c h_{ca} -\frac{1}{2} \p^2 h_{ab}  \right) \\
 \mathcal{H}^{\phi(2)}&=& \frac{p_\phi^2}{2\abar^{3}} +  \frac{\abar}{2} e^{ab} \p_a\phi\p_b\phi  + \frac{\abar^3}{2}\ V''(\bar{\phi})\ \phi^2 + \abar \left(-\frac{\bar{p}_\phi}{2\abar^6} \ p_\phi   + \frac{1}{2} V'(\bar{\phi})\ \phi \right)h  \nn\\
 &&    + \frac{\bar{p}_\phi^2}{8\abar^7}\left( h_{ab}h^{ab} + \frac{1}{2} h^2\right)   - \frac{1}{4\abar } V(\bar{\phi}) \left(h_{ab}h^{ab} -\frac{1}{2} h^2 \right)   \\
C_{a}^{(1)}&=& - 2\abar^2 \p^b p_{ab}    -\frac{\pbar}{3\abar}\left( \p^ch_{ac} -\frac{1}{2} \p_a h\right)      + \bar{p}_{\phi} \p_a \phi.
\ee
All indices in these equations are raised and lowered by the Euclidean metric  $e_{ab}$; $\p^2 = e^{ab}\p_a\p_b$, $h=h_{ab}e^{ab}$, and $p= p^{ab}e_{ab}$. The derivation of these expressions appears in Appendix \ref{HA}.

\subsection{Linearized theory in momentum space} 
   
We next write the action for the perturbations and the shift in spatial Fourier modes,  as this significantly simplifies the remaining analysis. We set 
\be 
\label{eq:sfs}
 h_{ab} (t,\vec{x}) &=& \int{d^{3}k\left[e^{i\vec{k}.\vec{x}}M_{ab}^I h_I(t,\vec{k})\right]}, \\
p^{ab} (t,\vec{x}) &=& \int{d^{3}k\left[e^{i\vec{k}.\vec{x}}M^{ab}_Ip^I(t,\vec{k})\right]},\\
\phi(t,\vec{x})&=&\int{d^{3}k\left[e^{i\vec{k}.\vec{x}}\tilde{\phi}(t,\vec{k})\right]},\\
p_\phi(t,\vec{x})&=&\int{d^{3}k\left[e^{i\vec{k}.\vec{x}}\tilde{p}_{\phi}(t,\vec{k})\right]},\\
\xi^{a} (t,\vec{x}) &=& \int{d^{3}k\left[e^{i\vec{k}.\vec{x}}\tilde{\xi}^{a}(t,\vec{k})\right]}.
 \ee
 Here the matrices $M_{ab}^I, \ I=1\cdots 6$ (to be defined below) form a time independent basis for $3\times 3$ symmetric matrices that give a decomposition of the gravitational  phase space variables into the canonical set $(h^I,p_I)$.  The matrices $M^I$  must satisfy the orthogonality condition
 \be
   \text{Tr} (M^IM^J) = M_{ab}^I M^{J ab} = \delta^{I J}, \label{orthogM}
 \ee
 to ensure that  the symplectic structure is preserved when the canonical action for perturbations \eq{pert-action} is written in $k-$space, i.e.
 \be
 \label{hpbasis}
 \int d^3x dt \ p^{ab}\dot{h}_{ab} \longrightarrow \int d^3k dt\  p_I\dot{h}^I.  
 \ee

 A suitable matrix basis that fulfills this requirement may be constructed using the unit mode vector and two unit orthogonal vectors in the plane transverse to $k^a$
 \be
 \eps_3^a\equiv  k^a/|k|, \ \  \eps_1^a,\ \ \epsilon_2^a.
 \ee
 Since we would like to characterize the matrices $M$ as having defined helicity with respect to rotations about the $k^a$ axis,  we replace
 $\eps_1^a, \eps_2^a$ with the eigenvectors of the rotation matrix $J_\theta$ about the $k^a$ axis. These are $\eps_\pm^a = (\eps_1^a \pm i\eps_2^a)/\sqrt{2}$, and satisfy  $J_\theta \eps_\pm =  e^{\pm i\theta} \eps_\pm$ , $J_\theta \eps_3 = \eps_3$, and  $e_{ab}\eps_-^a\eps_+^b=1$ and $e_{ab}\eps_\pm^a\eps_\pm^b=0$. Using the set $(\eps_3,\eps_\pm)$, the  Euclidean metric may be written as
 $e^{ab} = 2\eps_+^{(a}\eps_-^{b)} + \eps_3^a\eps_3^b$. 
 
  The six matrices $M^I$ are  constructed from the elements 
  \be
  \eps_3^a\eps_3^b,\ \ \eps_-^{(a}\eps_+^{b)}, \ \  \eps_3^{(a}\eps_\pm^{b)},   \ \ \eps_\pm^a \eps_\pm^b.
  \ee  
 Under $J_\theta$, the first two transform as scalars, the second two as vectors, and the last two as tensors. However as they stand, these do not satisfy the desired orthogonality conditions  \eq{orthogM}. This is achieved by the  following linear combinations:
 \be
M_{1}^{ab} &=& \frac{1}{\sqrt{3}}\ e^{ab},\\
M_{2}^{ab} &=& \sqrt{\frac{3}{2}} \left( \eps_{3}^{a}\eps_{3}^{b} - \frac{1}{3}  e^{ab} \right) ,\\
M_{3}^{ab} &=& \frac{i}{\sqrt{2}}\left(\eps_{-}^{a}\eps_{-}^{b}  - \eps_{+}^{a}\eps_{+}^{b}\right),\\
M_{4}^{ab} &=& \frac{1}{\sqrt{2}}\left(\eps_{-}^{a}\eps_{-}^{b}  + \eps_{+}^{a}\eps_{+}^{b}\right),\\
M_{5}^{ab} &=& i\left(\eps_{-}^{(a}\eps_{3}^{b)}  - \eps_{+}^{(a}\eps_{3}^{b)}\right),\\
M_{6}^{ab} &=& \eps_{-}^{(a}\eps_{3}^{b)}  + \eps_{+}^{(a}\eps_{3}^{b)}, 
\ee
 where again the first pair transform as scalars, the next pair as tensors, and the last pair as vectors. Let us also note a few other properties of these matrices: 
 \be
 e^{ab}M_{ab}^I &=&0, \ \ I=2\cdots 6; \nn\\
 k^aM_{ab}^I &=& 0,   \ \  I=3,4; \nn \\
 k^ak^bM_{ab}^I &=& 0, \ \ I = 5,6. 
 \ee
 Thus in the decomposition of  the Fourier transform of the metric perturbation $\tilde{h}_{ab}(k,t) \equiv M_{ab}^I h_I(k,t)$,  $h_1,h_2$ are the scalar modes, $h_3,h_4$ are the transverse traceless tensor modes, and $h_5,h_6$ are the transverse vector modes. The same properties hold  for the momenta $p_I$ conjugate to $h^I$. The shift perturbation may also be decomposed into longitudinal and transverse components:
 \be
 \tilde{\xi}^{a}(t,\vec{k}) &= \xi_{1}(t,\vec{k})\eps_{1}^{a} + \xi_{2}(t,\vec{k})\eps_{2}^{a} + \xi_{||}(t,\vec{k})\eps_{3}^{a}.
 \ee 
 
In summary, so far we have decomposed the perturbations $h_{ab}(x,t), p^{ab}(x,t)$ into longitudinal and transverse Fourier modes
$h^I(k,t),p_I(k,t), \ I=1\cdots 6$, with well defined  physical properties, and a related expansion for $\xi^a$. (The scalar field perturbation of course does not require any decomposition.) We now write the canonical action in $k-$space using this decomposition.
 
 \subsection{Canonical action in momentum space} 
As is standard in field theory, writing an action in momentum space using  \eq{eq:sfs} requires field redefinitions after implementing the reality conditions such as $\tilde{h}_{ab}^*(t,k) =  \tilde{h}_{ab}(t,-k)$. One way to do this is to write $\tilde{h}_{ab}(k,t) =  \tilde{h}_{ab}^R(k,t)+ i\tilde{h}_{ab}^I(k,t)$, impose the reality condition, restrict the action to be over independent modes, and then redefine modes to give an action with integration over all $k$.  Following these steps, and using the decompositions  
\be
\tilde{h}_{ab}(k,t) = M_{abI}h^I(k,t),\ \ \  \tilde{p}^{ab}(k,t) = M^{abI}p_I(k,t), \quad  i=1\cdots 6,
\ee
gives the $k-$space action 
\be
S^{(2)}=\int{dt d^{3}k\left[\dot{h}^Ip_I + \dot{\tilde{\phi}} \tilde{p}_\phi - \tilde{H}^{(2)} - i\tilde{\xi}^{a}\tilde{C}_a^{(1)}\right]},
\label{k-act}
\ee  
where $\tilde{H}^{(2)}=\mathcal{\tilde{H}}^{G(2)}+\mathcal{\tilde{H}}^{\phi(2)}$, and
 \begin{subequations}
\begin{align}
\label{hgk}
\tilde{\mathcal{H}}^{G(2)}=\ &\bar{a}\left(p^{I}p_{I}-\frac{3}{2}p_{1}^{2}\right)+ 
\frac{1}{\abar}\left(\frac{\bar{p}}{6\bar{a}}\right)\left(p^{I}h_{I}-\frac{3}{2}h_{1}p_{1}\right)
+\frac{1}{8\abar^3}\left(\frac{\bar{p}}{6\abar}\right)^{2}\left(5h_{I}h^{I}-\frac{9}{2}h_{1}^{2}\right)\nn\\
&  - \frac{k^2}{6\abar^3} \left[ \left(h_1-\frac{h_2}{\sqrt{2}}\right)^2  -\frac{3}{2} \left( h_3^2+h_4^2 \right)   \right],\\ 
\label{hmk}
\tilde{\mathcal{H}}^{\phi(2)}=\ & \frac{\tilde{p}_{\phi}^{2}}{2\bar{a}^{3}}   +\frac{\bar{a}}{2}k^{2}\tilde{\phi}^{2}  +\frac{\bar{a}^{3}}{2}V''(\bar{\phi})\tilde{\phi}^{2}\nn\\
& + \sqrt{3}\abar \left( -\frac{ \bar{p}_{\phi}}{2\bar{a}^6} \tilde{p}_{\phi}  +\frac{1}{2}V'(\bar{\phi}) \tilde{\phi}\right)h_1 \nn\\
& + \frac{\pbar_\phi^2}{8\bar{a}^{7}}\left(h_{I}h^{I}+\frac{3}{2}h_{1}^{2}\right) -\frac{V(\bar{\phi})}{4\bar{a}}\left(h_{I}h^{I}-\frac{3}{2}h_{1}^{2}\right).
\end{align}
\end{subequations}
The linearized diffeomorphism constraint in momentum space $\tilde{C}_{a}^{(1)}=0$ is
 \be
\tilde{C}_a^{(1)}&=&\tilde{C}_{a}^{G}+\tilde{C}_{a}^{\phi}\nn\\
&=&-2\bar{a}^{2}k^{b}M_{ab}^{I}p_{I}-\frac{\bar{p}}{3\bar{a}}\left(k^{c}M^{I}_{ac}h_{I}-\frac{\sqrt{3}k_{a}}{2}h_{1}\right)+\bar{p}_{\phi}k_{a}\tilde{\phi}.
\ee
A further expansion of $\tilde{C}_a^{(1)}$ using the properties of the $M^I$ basis reveals its longitudinal and transverse components:
\begin{subequations}
\begin{align}
\tilde{C}_a^{(1)}  = &\ k\left[  -\frac{2\abar^2}{\sqrt{3}}(p_{1}+\sqrt{2}p_{2}) + \frac{\pbar}{6\sqrt{3}\abar}(h_{1}-2\sqrt{2}h_{2})+\bar{p}_\phi \tilde{\phi}\right]\eps_{3a}\nn \\
& -\sqrt{2}k \left[\abar^2 p_6 + \left(\frac{\pbar}{6\abar}\right) h_6 \right] \eps_{1a} 
 -\sqrt{2}k \left[\abar^2 p_5 + \left(\frac{\pbar}{6\abar}\right) h_5 \right] \eps_{2a}.
\label{diff-comp}
\end{align}
\end{subequations}
 Similarly, the gravitational Hamiltonian may be written as a sum of scalar $(h_1,h_2)$, tensor $(h_3,h_4)$, and vector $(h_5,h_6)$ components, and their canonical momenta: 
\be
\tilde{H}^{G(2)} = H^S+H^V+H^T,\label{svt}
\ee 
\begin{subequations}
\begin{align}
\label{HS}
H^S  &=    \abar\left(p_2^2 -\frac{1}{2} p_1^2  \right) + \frac{1}{\abar} \left(\frac{\pbar}{6\abar}\right) \left(  h_2p_2 -\frac{1}{2} h_1p_1\right)  
  + \frac{1}{8\abar^3} \left(\frac{\pbar}{6\abar}\right)^2\left(\frac{1}{2} h_1^2 + 5h_2^2 \right) \nn\\ &\quad- \frac{1}{6\abar} \left(\frac{k}{\abar}\right)^2\left( h_1-\frac{1}{\sqrt{2}}h_2\right)^2 , \\
  \label{HV}
 H^V &=   \abar\left(p_5^2 + p_6^2\right)  +  \frac{1}{\abar} \left(\frac{\pbar}{6\abar}\right)  \left(p_5h_5+p_6h_6 \right) 
 +\frac{5}{8\abar^3} \left(\frac{\pbar}{6\abar}\right)^2   \left(h_5^2 + h_6^2 \right),\\
 \label{HT}
 H^T &=  \abar\left(p_3^2 + p_4^2\right) +  \frac{1}{\abar} \left(\frac{\pbar}{6\abar}\right) \left(p_3h_3+p_4h_4\right) 
 +\frac{1}{4\abar}\left[ \frac{5}{2\abar^2} \left(\frac{\pbar}{6\abar}\right)^2 +\left(\frac{k}{\abar}\right)^2\right] \left(h_3^2 + h_4^2 \right).
\end{align}
\end{subequations}
This shows that  only the scalar canonical pairs $(h_1,p_1)$ and $(h_2,p_2)$ interact with each other, while all the other pairs are mutually decoupled.  Denoting the longitudinal and transverse components of the diffeomorphism constraint by $C_{\parallel}$ and $C_{\perp}$, we  note also that  
\be
\quad \{H^S, C_\perp \} =0, \quad   \{H^V, C_\parallel \} =0, \quad \{ H^T, C_a \} = 0.   
\ee
Thus the graviton modes are diffeomorphism invariant  (to this order). All propagating modes appear with a factor $k^2$ so the vector modes are non-propagating; the last term in $H^S$ is the curvature perturbation up to an overall factor.

\subsection{Partial gauge fixing: removal of vector modes}

At this stage it is useful to perform a gauge fixing to remove the  vector modes. This involves imposing canonical gauge conditions on these modes and solving strongly the corresponding diffeomorphism constraint  components. The above decomposition reveals the convenient choice
\be
 h_5 = h_6 =0. \label{vec-gauge}
\ee
These are second class with the components $C_\perp$, 
\be
\{h_5, C_\perp\} = \{h_6, C_\perp \} = \sqrt{2}k\abar^2, 
\ee
unless $\abar=0$ or $k=0$.  Since we are interested in propagating modes (where the diffeomorphism constraint  is not identically zero),  and in regions far from a potential singularity, these gauge choices are sufficient. $C_\perp=0$ is then solved by setting $p_5=p_6=0$. 

The resulting  $\tilde{H}^{G(2)}$ is now 
\be
\tilde{H}^{G(2)} = H^S + H^T,
\ee
and the second order scalar field Hamiltonian becomes
\be
\label{hmk2}
\tilde{\mathcal{H}}^{\phi(2)} &=& \frac{\tp_{\phi}^{2}}{2\bar{a}^{3}}   +\frac{\bar{a}}{2}k^{2}\tphi^{2}  +\frac{\bar{a}^{3}}{2}V''(\bar{\phi})\tphi^{2}\nn\\
& &+ \sqrt{3}\abar\left[ -\left(\frac{\pbar_\phi}{2\abar^6}\right)  \tp_\phi  + \frac{1}{2}V'(\bar{\phi})\tphi  \right]h_1 \nn\\ 
&& + \frac{\pbar_\phi^2}{8\bar{a}^{7}}\left(h_{I}h^{I}+\frac{3}{2}h_{1}^{2}\right) -\frac{V(\bar{\phi})}{4\bar{a}}\left(h_{I}h^{I}-\frac{3}{2}h_{1}^{2}\right),
\ee   
where the sums $h_Ih^I$ in the last line now excludes the vectors modes $h_5,h_6$. The first  term is the standard hamiltonian of the scalar field  perturbation $(\tphi,\tp_\phi)$ on the $(\abar,\bar{\phi})$ homogeneous background; the second term contains the coupling of the scalar field perturbation  to the metric scalar mode $h_1$;  the last term is a potential for the graviton and metric-scalar modes.  

The diffeomorphism constraint is reduced to only  its longitudinal component
\be
\label{sc-diff}
 \tilde{C}_\parallel \equiv  - 2\abar^2 (p_{1}+\sqrt{2}p_{2}) + \left(\frac{\pbar}{6 \abar}\right)(h_{1}-2\sqrt{2}h_{2})+\sqrt{3} \bar{p}_\phi \tphi =0.
\ee
In summary, the gauge fixing (\ref{vec-gauge}) leaves a simpler system for the the remaining degrees of freedom: the metric scalar modes $(h_1,h_2)$,  graviton modes $(h_3,h_4)$, and the scalar field mode $\tphi$.  

\subsection{Graviton equation}
The graviton part of the second order canonical action is
\be
S^{g} \equiv \int dt d^3k \left[ p^I\dot{h}_I  - H^g \right], \ \ I=3,4,
\ee 
where $H^g$ is the sum of $H^T$ in \eq{HT} and the graviton parts of  $\mathcal{H}^{\phi(2)}$ in \eq{hmk}. For comparison with covariant perturbation theory, where the expansion 
$ q_{ab} = \abar^2(t)\left( e_{ab} +  h_{ab}\right)$ is used, let us make the   transformation
\be
h_{I} \longrightarrow \abar^2 h_I, \quad  p_I  \longrightarrow \abar^{-2}p_I. 
\ee
With this,  the symplectic term  transforms to 
\be
\dot{h}_Ip^I \longrightarrow \dot{h}_Ip^I + 2 \left(\frac{\dot{\abar}}{\abar}\right) h_Ip^I =   \dot{h}_Ip^I - \frac{\pbar}{6\abar^2}h^Ip_I,
\ee
where the last step uses the e.o.m. of the background. Therefore $H^g$ transforms to 
\be
H^g &=&   \frac{1}{\abar^3}\left(p_3^2 + p_4^2\right) +   \left(\frac{\pbar}{3\abar^2}\right) \left(p_3h_3+p_4h_4\right) \nn\\
 &&    +  \frac{\abar^3}{4}\left[  \frac{\pbar_\phi^2}{2\abar^6} - V(\bar{\phi}) + \frac{5}{2} \left(\frac{\pbar}{6\abar^2}\right)^2 + \left(\frac{k}{\abar}\right)^2      \right]  \left(h_3^2 + h_4^2\right). \label{g-H}
\ee
Although this expression for $H^g$ looks involved, it is readily verified that the canonical equations of motion
\be
 \dot{h}_I = \{ h_I, H^g\}, \quad  \dot{p}_I = \{ p_I, H^g\},
\ee
together with the equations (\ref{eq:zeroeqns}) of the background $(\abar,\pbar)$,  leads to the standard wave equation
\be
\ddot{h}_{I}+3\left( \frac{\dot{\abar}}{\abar}\right)\dot{h}_{I}+\frac{k^{2}}{a^{2}}h_{I}=0, \qquad I=3,4.
\ee
Thus  the graviton mode equation is unchanged in the canonical dust-time gauge. The calculation leading to this has some non-trivial steps (see Appendix \ref{gravitonA}). 
 
\section{Scalar modes}

We have so far seen that the dust-time physical Hamiltonian in momentum space, in the time independent matrix basis $M$, provides a relatively simple way to analyze cosmological perturbations. Specifically we showed from a canonical perspective how the vector perturbations  are removed, and the graviton equation remains unchanged. 

We now turn to the remaining degrees of freedom  $(h_1,h_2, \tphi)$, with dynamics described by $H^S$ \eq{HS} and   $\tilde{\mathcal{H}}^{\phi(2)}$  \eq{hmk2},  
\be
H^{S\phi}  &\equiv&  \abar\left(p_2^2 -\frac{1}{2} p_1^2  \right) + \frac{1}{\abar} \left(\frac{\pbar}{6\abar}\right) \left(  h_2p_2 -\frac{1}{2} h_1p_1\right)  
  + \frac{1}{8\abar^3} \left(\frac{\pbar}{6\abar}\right)^2\left(\frac{1}{2} h_1^2 + 5h_2^2 \right) \nn\\ 
&&  - \frac{1}{6\abar} \left(\frac{k}{\abar}\right)^2\left( h_1-\frac{1}{\sqrt{2}}h_2\right)^2   + \frac{\pbar_\phi^2}{8\bar{a}^{7}}\left(\frac{5}{2}h_1^2 +h_2^2 \right) + \frac{V(\bar{\phi})}{4\bar{a}}\left(\frac{1}{2}h_{1}^{2}-h_2^2\right)\nn\\
&& + \frac{\tp_{\phi}^{2}}{2\bar{a}^{3}}   +\frac{\bar{a}}{2}k^{2}\tphi^{2}  +\frac{\bar{a}^{3}}{2}V''(\bar{\phi})\tphi^{2} 
  +\sqrt{3}\abar\left[ -\left(\frac{\pbar_\phi}{2\abar^6}\right)  \tp_\phi  + \frac{1}{2}V'(\bar{\phi})\tphi  \right]h_1, 
  \label{HSphi}
  \ee  
  subject to the remaining diffeomorphism constraint $C_\parallel$ \eq{sc-diff}.  
  
 The Hamiltonian $H^{S\phi}$  is of the form of $h_S(h_i,p_i) + h_\phi(\tphi,\tp_\phi) + h_\text{Int}(\tphi,\tp_\phi,h_1)$. It is  notable that the scalar field perturbation $\tphi$ interacts with only the metric-scalar mode $h_1$ in the last term.  The constraint  $C_\parallel$  depends on the remaining phase space variables, and is also explicitly time dependent through the background solution $(\abar,\pbar,\pbar_\phi)$; it is therefore useful to check that it is remains first class, i.e.
 \be
 \dot{\tilde{C}}_\parallel = \left\{\tilde{C}_\parallel, \int d^3k \ H^{S\phi} \right\}  + \frac{\p \tilde{C}_\parallel}{\p t}   =0.
 \ee 
This is indeed the case (see Appendix \ref{diffA}).  

At this stage we have one first class constraint $C_\parallel$ and  three configuration variables $h_1,h_2,\phi$. Therefore there are two physical configuration degrees of freedom in the metric perturbation (in addition to the two graviton modes we have already discussed).   We recall that this is unlike the standard cosmological perturbation theory where the  starting point has only the  metric and scalar field perturbations; in the model we are studying, there is also the dust field, which  was fixed as the time coordinate, thereby leaving an additional physical configuration variable in the metric perturbation. We now turn to identifying two physical diffeomorphism invariant variables and their conjugate momenta.  These satisfy 
\be
\{ {\cal O}, \tilde{C}_\parallel\} =0.  \label{obs}
\ee

\subsection{Diffeomorphism invariant observables}

 For linear perturbation we are interested in observables ${\cal O}$ defined by \eq{obs}   that are linear in the phase space variables $(h_1,h_2,p_1,p_2,\tphi,\tp_\phi)$. There are many choices.  We are interested in diffeomorphism invariant canonical pairs and an expression for the physical Hamiltonian \eq{HSphi} in terms of such pairs. Let us note that $\tphi$ is already invariant since $\tp_\phi$ does not appear in $\tilde{C}_\parallel$. A few  other elementary  ones are 
 \be
  && H\equiv h_{1}-\frac{h_2}{\sqrt{2}}, \quad\quad\quad \quad \ \  \ \ P\equiv p_1 + \frac{p_2}{2\sqrt{2}}, \\
   && A_1\equiv h_1 - \left( \frac{12\abar^3}{\pbar} \right)p_1, \quad \ \ A_2\equiv h_2 + \left(\frac{6\abar^3}{\pbar} \right)   p_2, \\
   &&B_1= h_1 -\left(\frac{2\abar^2}{\sqrt{3}\pbar_\phi} \right)\tp_\phi, \quad B_2 = h_2 - \left(\frac{2\sqrt{2} \abar^2}{3\pbar_\phi} \right) \tp_\phi.
\ee
These may be used to construct invariant canonical pairs  by taking linear combinations with coefficients that are functions  of the background solution. 

The first of these observables $H$ is proportional to the Ricci curvature $R^{(3)}$ of the spatial slice. To see this we note that  to linear order 
\be
  R^{(3)}=\frac{1}{\abar^{4}}\left(\p_a\p_b h^{ab} - \p^2 h\right).
\ee
In the $M$ basis in momentum space, this  becomes
\be
\tilde{R}^{(3)}=4\left(\frac{k}{\abar}\right)^2\left[\frac{1}{2\sqrt{3}\abar^{2}}\left(h_{1}-\frac{h_{2}}{\sqrt{2}}\right)\right]\equiv 
-4\left(\frac{k}{\abar}\right)^2\psi;
\label{psi}
\ee
The $\psi$ in the last term defines the curvature perturbation used in the  covariant theory. It is readily verified that a momentum conjugate to $\psi$ is  
\be
P_\psi \equiv -\frac{8\abar^2}{\sqrt{3}} \left(p_{1}+\frac{p_{2}}{2\sqrt{2}}\right).
\ee
This satisfies 
\be
  \quad \{P_\psi, \tilde{C}_\parallel\} =0, \quad  \{\psi, P_\psi\} =1.
\ee
  A second canonical pair is found by noting that the scalar field perturbation $\phi$ is diffeomorphism invariant,
 \be
 \{ \phi, \tilde{C}_\parallel\} =0. 
 \ee
 For notational convenient we define $\gamma \equiv \tphi$. A diffeomorphism  invariant  variable canonically conjugate to $\gamma$  is 
\be
P_\gamma = \tp_\phi + 2\sqrt{3}\ \frac{\abar\pbar_\phi}{ \pbar}   \left(p_{1}+\sqrt{2}p_{2}\right),
\ee
and this  satisfies
\be
\{P_\gamma, \tilde{C}_\parallel\} =  \{P_\gamma, \psi \} =\{P_\gamma, P_\psi \} =0, \quad \{\gamma,P_\gamma\}=1.
\ee 
(We note that $\tp_\phi$ is canonically conjugate to $\tphi$, but it is not gauge invariant, hence the need to define an alternative conjugate momentum that is gauge invariant.) 
Although the Hamiltonian \eq{HSphi} may be written down in terms of these variables, it is more convenient to use a different set that is useful to make contact with the conventional perturbation theory without the dust field. For this reason we select the following diffeomorphism invariant canonical pairs. The first pair is 
\be 
\mathcal{R}&=& \psi -\left(\frac{\abar\pbar}{12\pbar_{\phi}}\right)\tphi \\
P_{\mathcal{R}}&=& \left( \frac{48k^2\abar^3}{\pbar}  \right)\psi + \sqrt{\frac{2}{3}} \left(\frac{\pbar}{\abar} \right) A_2,
\label{RPR}
\ee
and the second pair is 
\be
 \chi &=& \left(\frac{\abar^3}{\pbar_\phi} \right) \tphi, \\
 P_\chi &=& 4 \abar k^2 \psi  + \left(\frac{\sqrt{3}\pbar}{18\abar}\right)\left(\frac{\dot{\pbar}_\phi}{\pbar_\phi} - 3\bar{H} \right) A_1  - \sqrt{\frac{2}{3}} \left(\frac{\pbar \dot{\pbar}_\phi}{3\abar \pbar_\phi} \right) A_2 - \left(\frac{\sqrt{3} \pbar_\phi^2 }{2\abar^5} \right)B_1.  
 \ee
These satisfy 
\be
\{{\cal R},P_{\mathcal{R}}\} =  \{\chi,P_\chi \} = 1,  \quad \{P_{\mathcal{R}}, P_\chi \} =\{P_{\mathcal{R}}, \chi \} = \{P_\chi, {\cal R} \}=\{ {\cal R},\chi\}=0.
\ee

We can now write the Hamiltonian \eq{HSphi}  in terms of these canonical variables. Before doing this it is convenient to fix a gauge and solve the diffeomorphism constraint $C_\parallel =0$;  since the variables are diffeomorphism invariant, their values would of course be unaffected.  We choose the gauge 
\be
h_{1}=0. 
\ee
This choice  removes the interaction of $h_1$ and $\phi$ in the Hamiltonian  (\ref{HSphi}), thereby simplifying it considerably.   It is second class with $C_\parallel$:
\be
 \{h_1, C_\parallel \} = -2\abar^2,
\ee
unless $\abar =0$. Setting $h_1=0$ and solving the diffeomorphism constraint for $p_1$, 
\begin{equation}
p_{1} = -\sqrt{2}\left(p_{2}+\frac{\pbar}{6\abar^3}h_2\right)+\frac{\sqrt{3}p_{\bar{\phi}}}{2\abar^2} \tphi,
\end{equation}
gives the fully reduced theory for the  gauge invariant pairs $({\cal R}, P_{\mathcal{R}})$ and $(\chi,P_\chi)$.  The final action is
\begin{equation}
\label{eq:simpleaction}
S^{(2)}_{GF}\equiv\int{dt d^{3}k\left[\dot{{\cal R}}P_{\mathcal{R}}+\dot{\chi}P_{\chi} - H^{(2)} \right]},
\end{equation}
where the $k-$space Hamiltonian density takes the remarkably simple form
\be 
H^{(2)} = \frac{1}{2 \abar}\left[\frac{1}{z^2} P_{\mathcal{R}}^2 +  k^2 \left(z{\cal R}\right)^2   \right]  
 +\left( \frac{\bar{a}^{3}}{2\bar{p}_{\phi}^{2}}\right)P_{\chi}^{2}- \left(\frac{\bar{a}\bar{p}}{12\bar{p}_{\phi}^{2}}\right)P_{\mathcal{R}}P_{\chi},
\label{dustH}
\ee
with
 \be
z= -  \frac{12\pbar_\phi}{\pbar}. 
 \ee
 The equations of motion following from this Hamiltonian are 
 \be
 \dot{\cal R} &=&\left( \frac{1}{\abar z^2}\right) P_{\cal R} + \left( \frac{\abar}{z\pbar_\phi}\right)P_\chi,\\
 \dot{P}_{\cal R} &=&- \left(\frac{k^2z^2}{\abar}\right) {\cal R},\\
 \dot{\chi} &=& \left( \frac{\bar{a}^{3}}{\bar{p}_{\phi}^{2}}\right) P_\chi + \left(\frac{\bar{a}}{z\pbar_\phi}\right)P_{\mathcal{R}} = \left(\frac{1}{\bar{H}}\right)\dot{\cal R}\\
 \dot{P}_\chi &=&0 \implies P_\chi =C. 
 \ee
 These lead to the second order equations 
 \be
 &&\ddot{\cal R} + \left(\frac{\dot{\zeta}}{\zeta}\right)  \dot{\cal R} + \left(\frac{k^2}{\abar^2}\right) {\cal R} = C \bar{f}(t), 
 \label{dust1}\\
 &&\ddot{\chi}  +\left(\frac{\dot{\alpha}}{\alpha}   \right) \dot{\chi} = 
  \frac{1}{\bar{H}} \left(C\bar{f} - \frac{k^2}{\abar^2} {\cal R}\right),
 \label{dust2}
 \ee
 where $\zeta = \abar z^2$, $\alpha = \bar{H}\zeta$ and $\bar{f}(t)$ is the following function of background solution
 \be
 \bar{f} = \left(\frac{\abar }{z\pbar_\phi} \right)^\cdot  +  \left(\frac{\dot{\zeta}}{\zeta}\right)\left(\frac{\abar }{z\pbar_\phi} \right) 
 = \frac{\dot{\bar{H}}}{\abar^3\dot{\bar{\phi}}}.
 \ee
 Thus the equation for ${\cal R}$ resembles that obtained in the usual cosmological perturbation theory, but now has a forcing term that is a function $\bar{f}$ of the background fields and $P_\chi=C$; for the choice $C=0$ this equation is the same as that in usual cosmology.  The equation for $\chi$ on the other hand is  ultra-local because there is no term in it of the form $k^2\chi$,  which would indicate the presence of spatial derivatives of $\chi$;  $k$ dependence of $\chi$ therefore arises solely from the source term of \eq{dust2}. This is not surprising since we would not expect a second propagating degree of  freedom starting with a theory containing pressureless dust. Indeed this is also what is obtained for perturbation theory on flat spacetime \cite{Ali_2016}. As a final comment in these equations we note that (\ref{dust2}) may be rewritten using the variable 
 \be
 \tilde{\chi} \equiv \chi - \frac{{\cal R}}{\bar{H}} ,
 \ee
 leading to 
 \be
 \ddot{\tilde{\chi}} + \left( \frac{\dot{\alpha}}{\alpha}\right)  \dot{\tilde{\chi}} = \frac{1}{\alpha}\frac{d}{dt}\left[\left(\zeta \frac{\dot{\bar{H}}}{\bar{H}} \right) {\cal R} \right],
 \ee
 which removes the $k^2$ term on the r.h.s. of (\ref{dust2}). This shows the ultralocality of $\tilde{\chi}$ due to the absence of the spatial derivative propagation term $k^2\tilde{\chi}$ -- the same reasoning as for $\chi$. 
  
  Let us summarize the results so far. We started with the theory of GR coupled to dust and a scalar field. This theory has four physical field degrees of freedom, of which 2 are  gravitational. The Hamiltonian perturbation analysis we presented therefore must also have the same number. By fixing the dust time gauge, one of the these four degrees of freedom manifests itself in the metric. Thus, after identifying the two graviton modes, we are left with an additional scalar mode, which as we have seen turns out to be ultralocal.  
   
 \section{Comparison with perturbation theory without dust} 

 It is useful to compare the dust time perturbation theory we have developed above with a similar hamiltonian treatment of standard perturbation theory. This begins with the ADM hamiltonian  action of GR coupled to only a scalar field. This is eqn. (\ref{can-act}) with $T=P_T=0$.  Expansion of this action about a homogeneous and isotropic background solution  is of the form (\ref{eq:genans}), with the additional expansion of the lapse function
 \be
 N(x,t) = \bar{N}(t) + \delta N(x,t),
 \ee
 where we have taken $\bar{N}(t)$ as the lapse function of the background. The second order action changes from (\ref{pert-action}) to  
 \be
 {\cal S}^{(2)} \equiv \int d^3xdt \left[\dot{h}_{ab}p^{ab}+\dot{\phi}p_\phi - \delta N {\cal H}^{(1)}-  \bar{N}(t) {\cal H}^{(2)} - \xi^{a}C_{a}^{(1)}\right],
 \label{nodustS2}
 \ee
where $ {\cal H}^{(2)}$ and $C_{a}^{(1)}$ are exactly as given in (\ref{hgr2}), and 
\begin{equation}
\begin{split}
\mathcal{H}^{(1)}&=-\frac{1}{\bar{a}}\Bigg\{\left(\frac{\bar{p}}{6\bar{a}}\right)\left[\frac{1}{4}\left(\frac{\bar{p}}{6\bar{a}}\right)h+\bar{a}^{2}p\right]+\partial_{a}\partial_{b}h^{ab}-\partial^{2}h\Bigg\}\\&+\frac{\bar{p}_{\phi}}{\bar{a}^{3}}\left[p_{\phi}-\frac{\bar{p}_{\phi}}{4\bar{a}^{2}}h\right]+\bar{a}\left[\bar{a}^{2}V'(\bar{\phi})\phi+\frac{V(\bar{\phi})}{2}h\right].
\end{split}
\end{equation} 
We recall  that $h=h_{ab}e^{ab}$ and $p=p^{ab}e_{ab}$. In the following we will take the background lapse $\bar{N}(t)=1$.  
 
 We see that this second order action has two constraints obtained by varying w.r.t. the lapse and shift perturbation $\delta N(x,t)$ and $\xi^a(x,t)$. The action also displays a non-vanishing Hamiltonian $ \bar{N}(t) {\cal H}^{(2)}$, where $\bar{N}(t)$ is a fixed background function that cannot be varied in the second order action; it is of course varied in the zeroth order action to give the background Hamiltonian constraint $\bar {\cal H}=0$.  Thus, in comparison to the dust time gauge theory,  we have the additional constraint $\mathcal{H}^{(1)}=0$.  In momentum space, in the basis ($h_I, p^I$) (\ref{hpbasis}), this is expanded as 
 \begin{subequations}
\begin{align}
\begin{split}
\tilde{\mathcal{H}}^{(1)}&=-\frac{1}{\abar}\Bigg\{\left(\frac{\bar{p}}{6\bar{a}}\right)\left[\frac{1}{4}\left(\frac{\bar{p}}{6\bar{a}}\right)\tilde{h}+\bar{a}^{2}\tilde{p}\right]-k_{a}k_{b}\tilde{h}^{ab}+k^{2}\tilde{h}\Bigg\}\\&\quad +\frac{\pbar_{\phi}}{\abar^{3}}\left[\tilde{p}_\phi-\frac{\pbar_\phi}{4\bar{a}^{2}}\tilde{h}\right]+\bar{a}\left[\bar{a}^{2}V'(\bar{\phi})\tilde{\phi}+\frac{V(\bar{\phi})}{2}\tilde{h}\right]
\end{split}\\
\begin{split}
&=-\frac{\sqrt{3}}{\bar{a}}\Bigg\{\left(\frac{\pbar}{6\bar{a}}\right)\left[\frac{1}{4}\left(\frac{\bar{p}}{6\bar{a}}\right)h_{1}+\bar{a}^{2}p_{1}\right]+\frac{\sqrt{2}k^{2}}{3}\left[\sqrt{2}h_{1}-h_{2}\right]\Bigg\}\\&\quad+\frac{\bar{p}_{\phi}}{\bar{a}^{3}}\left[\tp_{\phi}-\frac{\sqrt{3}\bar{p}_{\phi}}{4\bar{a}^{2}}h_{1}\right]+\bar{a}\left[\bar{a}^{2}V'(\bar{\phi})\tphi+\frac{\sqrt{3}V(\bar{\phi})}{2}h_{1}\right].
\end{split}
\end{align}
\end{subequations}
It is important to note that $\tilde{\mathcal{H}}^{(1)}$ is a function of only the scalar metric modes $h_1$ and $h_2$ and their conjugate momenta $p_1$ and $p_2$, in addition to scalar field perturbations $(\tphi,\tp_\phi)$ -- the graviton modes appear only in $\tilde{\mathcal{H}}^{(2)}$.

After solving the transverse parts of the diffeomorphism constraints and removing the vector modes as before, only the parallel component of the constraint $C_\parallel=0$ \eq{sc-diff} remains. The momentum space action for the scalar perturbations $(h_1,h_2,\phi)$ becomes
\be
 S^{S\phi} \equiv \int dt d^3k \left[p_1 \dot{h}_1  + p_2\dot{h}_2 + \tp_\phi \dot{\tphi}  - H^{S\phi}
 - \delta \tilde{N}\tilde{\mathcal{H}}^{(1)}- \tilde{\xi}_\parallel \tilde{C}_\parallel^{(1)}  \right]
 \label{Sphi-act}
 \ee
 where $H^{S\phi}$  is given in  (\ref{HSphi}), and $\delta \tilde{N}(k,t)$ is the lapse perturbation in momentum space. We now note that the constraints obtained by varying this action w.r.t. $\tilde{\xi}_\parallel$ and $\delta \tilde{N}(k,t)$ are first class. We have already verified that $C_\parallel$ is first class (Appendix \ref{diffA}). We also find that  
\be
\frac{d}{dt}\ \tilde{\mathcal{H}}^{(1)} = \{ \tilde{\mathcal{H}}^{(1)}, H^{S\phi}  \}  + \frac{\p}{\p t} \tilde{\mathcal{H}}^{(1)}= \tilde{C}_\parallel=0,
\ee
 and 
 \be
 \{\tilde{\mathcal{H}}^{(1)}, \tilde{C}_\parallel \} = -\bar{\cal H}=0,
 \ee
 where the last equality follows from the background hamiltonian constraint $\bar{\cal H}=0$; recall that this is the theory without dust. This is a satisfying structure demonstrating  explicitly  that  the second order perturbed system is first class.   It also shows that, of the three scalar perturbation modes $(h_1,h_2,\tphi)$, only one is a physical degree of freedom (due to the two constraints $\tilde{\mathcal{H}}^{(1)}=0$ and $\tilde{C}_\parallel=0$). We can now proceed to obtain gauge invariant observables, i.e. those that Poisson commute with   
 $\tilde{C}_\parallel$ and $ \tilde{\mathcal{H}}^{(1)}$.  We note that, unlike the case with dust,  only one canonical pair of gauge invariant variables is required (due to the presence of two constraints instead of one).  
 
 \subsection{Gauge invariant variables}
 
 Gauge invariant variables ${\cal O}$ must now satisfy 
\be
  \{{\cal O}, \tilde{\mathcal{H}}^{(1)}\} =  \{ {\cal O}, \tilde{C}_\parallel\} =0. 
\ee 
 
   We have already noted that the curvature  perturbation 
 \be
\psi=-\frac{1}{2\sqrt{3}\abar^{2}}\left(h_{1}-\frac{h_{2}}{\sqrt{2}}\right)
 \ee
  defined in \eq{psi}  satisfies $\left\{\psi, C_\parallel \right\}=0$. However
 \be
 \big\{\psi,\tilde{\mathcal{H}}^{(1)}\big\}=\frac{\pbar}{12\abar^2}\ne 0,
 \ee
 therefore $\psi$ is not invariant under the second constraint, and therefore not fully gauge invariant.  By noting that 
 \be
 \big\{\tphi,\tilde{\mathcal{H}}^{(1)}\big\}=\frac{\pbar_{\phi}}{\abar^3},
 \ee
 we observe that the linear combination 
 \be
 {\cal R} \equiv \psi - \left(\frac{\abar\pbar}{12 \pbar_\phi}\right)\ \tphi,
 \ee
  satisfies 
 \be
 \left\{{\cal R}, \tilde{\mathcal{H}}^{(1)}\right\}=0, \quad \left\{{\cal R}, \tilde{C}_\parallel\right\} =0. 
 \ee
 This  ${\cal R}$  is exactly the same variable we used for the dust case. We have now learned that it is also invariant under the transformation generated by $\mathcal{H}^{(1)}$.   Similarly we note that its conjugate momentum defined in
 \eq{RPR} satisfies 
 \be
 \left\{ {\cal R}, P_{\cal R}  \right\} =1,  \quad \left\{ P_{\cal R}, \tilde{\mathcal{H}}^{(1)} \right\}=0, \quad \left\{ P_{\cal R}, \tilde{C}_\parallel\right\} =0.
 \ee
Thus the canonically conjugate pair $({\cal R}, P_{\cal R})$ are fully gauge invariant to this order. 
 
 We note also that any scaled variables of the type $ ( g {\cal R}, P_{\cal R}/g)$, where 
$g= g(\abar,\pbar,\bar{\phi},\pbar_\phi)$ is an arbitrary function of the background variables, are also gauge invariant (since the fixed background does not participate in the Poisson bracket for the perturbations). The choice
\be
g = -\frac{12\pbar_\phi}{\pbar} \equiv z
\ee
 gives the  Mukhanov-Sasaki (MS) variable
 \be
 \nu \equiv  -\left(\frac{12\pbar_\phi}{\pbar} \right){\cal R}= \left(\frac{\abar\dot{\bar{\phi}}}{\bar{H}}\right) {\cal R} =\abar\left(\tphi + \frac{\dot{\bar{\phi}}}{\bar{H}}\ \psi\right),
 \label{MS}
 \ee
 where the second equality follows from the background equations \eq{eq:zeroeqns}.

 \subsection{Gauge fixed action}
 
 As the last step, we fix two gauges corresponding to the two first class constraints $\tilde{\mathcal{H}}^{(1)}=0$ and $\tilde{C}_\parallel=0$, and solve these constraints strongly to obtain the final canonical action from \eq{Sphi-act} for the remaining unconstrained gauge invariant physical degrees of freedom. The final action will be a functional of the canonical pair ${\cal R}, P_{\cal R}$. This may then be recast in terms of  the MS variable $\nu$ and its conjugate momentum $P_\nu$.
  
 We set the gauge conditions 
 \be
\tphi=0, \quad  h_1=0. 
 \ee
 These satisfy 
 \be
 \{ \tphi, \tilde{\mathcal{H}}^{(1)} \} = \frac{\pbar_\phi}{\abar^3};\quad \{ h_1, \tilde{C}_\parallel\} =  -2\abar^2,
 \ee
therefore the constraints and gauge conditions form  second class pairs. Solving the constraints for $p_1$ and $\tp_\phi$ gives
\be
p_1 &=&-\frac{\sqrt{2}}{\bar{a}^{2}}\left[\left(\frac{\bar{p}}{6\bar{a}}\right)h_{2}+\bar{a}^{2}p_{2}\right]\\
 \tp_\phi&=&-\frac{1}{\sqrt{6}\bar{p}_{\phi}}\left(\frac{\bar{p}^{2}}{6}h_{2}+2\bar{a}^{2}k^{2}h_{2}+\bar{a}^{3}\bar{p}p_{2}\right).
\ee
In this gauge,  the invariant variables  ${\cal R}$ and $P_{\cal R}$ become
\be
\mathcal{R}=\frac{1}{2\sqrt{6}\abar^2} \ h_2, \quad P_{\mathcal{R}}=\sqrt{\frac{2}{3}} \left( \frac{\pbar}{\abar} + \frac{12\abar}{\pbar}k^2   \right)\ h_{2}+2\sqrt{6}\abar^{2}\ p_{2}.
\ee
Substituting the gauge conditions and solutions of the constraints into the action \eq{Sphi-act}, and expressing variables in terms of ${\cal R}$ and $P_{\cal R}$,  gives  
\be
S^{S}_{GF} \equiv \int dt d^3k \left[ \dot{\mathcal{R}}P_{\mathcal{R}}-H^{S}_{GF}  \right],
\label{MSact}
\ee
where 
 \be
 H^S_{GF}= \frac{1}{2\abar} \left[  \frac{1}{z^2}  P_{\mathcal{R}}^{2} +   k^2\left(z\mathcal{R}\right)^2 \right].
 \label{MSH}
 \ee
 This is the same as the action for the dust-time case \eq{dustH}, but with $\chi=P_\chi=0$.  
 
 As the last step in comparison with standard perturbation theory, we derive from this action the MS equation. We noted the definition of the MS variable $\nu$ in \eq{MS}. The conjugate momentum is $P_\nu = P_{\cal R}/z$.
 The action \eq{MSact} transforms to
 \be
 S^{S}_{GF} = \int dtd^3k \left[P_\nu\dot{\nu} - H_\nu \right],
 \ee 
  with 
 \be
 H_\nu = \frac{1}{2\abar}\left( P_\nu^2 +  k^2\nu^2  \right)  + \frac{\dot{z}}{z} \nu P_\nu.
 \ee 
 This gives the equation of motion
 \be
 \ddot{\nu} + \bar{H}\dot{\nu} +\left( \frac{k^2}{\abar^2} - \frac{\ddot{z}}{z} - \bar{H} \frac{\dot{z}}{z} \right) \nu =0.
 \ee
 In conformal time $dt = \abar d\tau$ this becomes the familiar MS equation
 \be
 \nu'' + \left(k^2 -\frac{z''}{z}\right)\nu =0.
 \ee
 
 To summarize this section, we have seen that the gauge invariant canonical variables $({\cal R}, P_{\cal R})$ that we used in the dust-time setting are  also invariant under the local time transformation generated by the additional constraint $\tilde{\mathcal{H}}^{(1)}$. This is in fact why we used these for the dust-time case, rather than variables that are only invariant under the diffeomorphism constraint $\tilde{C}_\parallel$. There are many other possibilities for canonical pairs  invariant under only the latter, but these do not provide a direct connection with the standard perturbation theory.  
 
 \section{Summary and discussion}
  
  We presented the hamiltonian theory of cosmological perturbations for GR coupled to dust and a scalar field, in the dust time gauge. The analysis demonstrates the following features: (i) the graviton modes decouple from other degrees of freedom and their equations of motion are unchanged, (ii) the vector modes are removed by gauge fixing in the same way as for flat space perturbation theory \cite{Ali_2016}, (iii) there  remain two coupled scalar modes,  one of which (${\cal R}$) satisfies a wave equation with a source, and the other $(\chi)$ satisfies an ultra-local equation with a source dependent on $k$; these two equations generalize the usual perturbation equations.  
    
 We also applied the same Hamiltonian decomposition, using the canonical variables $(h_I,p^I)$  to the standard cosmological perturbation theory. This differs from the Hamiltonian formalism  presented in \cite{Langlois:1993} in several respects. These include our use of a scale factor independent basis for decomposing metric perturbations, a demonstration that the perturbed constraints are first class, a  calculation of the constraint  algebra, and finally a step-by-step application of the reduction to physical degrees of freedom using the Dirac procedure.  Thus our work provides a more detailed view of Hamiltonian perturbation theory for cosmology, in addition to its extension to the dust time gauge.
 
Our final equations in the dust time gauge \eq{dust1} and \eq{dust2}  lead ultimately to the MS equation with an external forcing term dependent on the background solution, and an additional ultra-local equation for the field $\chi$. These may have observational consequences which we intend to explore in future work. The special solution $P_\chi=0$  removes the source term,  and so leads to exactly the MS equations plus the equation for $\chi$. However this case contributes no additional energy density since the terms proportional to $P_\chi$ in the hamiltonian density \eq{dustH} vanish for this case. Therefore the general case $P_\chi\ne0$ is more interesting for exploring cosmological consequences. 
\medskip

{\bf Acknowledgements} We thank Marco de Cesare, Edward Wilson-Ewing, and Suprit Singh for discussions and comments on the manuscript.
This work was supported by the NSERC of Canada.

\appendix 
\section{Derivation of second order Hamiltonian}\label{HA}

Recall that the physical Hamiltonian density for general relativity consists of a curvature and kinetic part:
\begin{equation}
\label{eq:grham}
\mathcal{H}^{GR} = -\sqrt{q}R^{(3)} + \frac{\pi_{ab}\pi^{ab}}{\sqrt{q}} - \frac{\pi^{2}}{2\sqrt{q}}.
\end{equation}
We list the expansions of the different pieces. The metric and its inverse are:
\begin{subequations}
\begin{align}
\label{eq:metric}
q_{ab}&=\bar{a}^{2}e_{ab}+\epsilon h_{ab}\\
q^{ab}&=\frac{e^{ab}}{\bar{a}^{2}}-\epsilon\frac{h^{ab}}{\bar{a}^{4}}
\end{align}
\end{subequations}
where $\epsilon$ tracks the order in perturbation. We will first compute the determinant using the usual definition:
\begin{equation}
q=\frac{\varepsilon^{abc}\varepsilon^{def}}{3!}q_{ad}q_{be}q_{cf}
\end{equation}
where $\varepsilon^{abc}$ is the Levi-Civita symbol. We expand the metric as defined in equation \eqref{eq:metric} and follow the steps detailed below to obtain the metric determinant.
\begin{equation}
\begin{split}
q&=\frac{\varepsilon^{abc}\varepsilon^{def}}{3!}\left(\bar{a}^{2}e_{ad}+\epsilon h_{ad}\right)\left(\bar{a}^{2}e_{be}+\epsilon h_{be}\right)\left(\bar{a}^{2}e_{cf}+\epsilon h_{cf}\right)\\
&=\frac{\bar{a}^{6}}{3!}\varepsilon^{abc}\varepsilon^{def}e_{ad}e_{be}e_{cf}+\frac{\epsilon \bar{a}^{4}}{2}\varepsilon^{abc}\varepsilon^{def}e_{ad}e_{be}h_{cf}+\frac{\epsilon^{2}\bar{a}^{2}}{2}\varepsilon^{abc}\varepsilon^{def}e_{ad}h_{be}h_{cf}\\
&=\frac{\bar{a}^{6}}{3!}\varepsilon_{def}\varepsilon^{def}+\frac{\epsilon \bar{a}^{4}}{2}\varepsilon_{de}^{\ \ c}\varepsilon^{def}h_{cf}+\frac{\epsilon \bar{a}^{2}}{2}\varepsilon_{d}^{\ bc}\varepsilon^{def}h_{be}h_{cf}\\
&=\bar{a}^{6}+\epsilon \bar{a}^{4} e^{cf}h_{cf} + \frac{\epsilon^{2}\bar{a}^{2}}{2}\left(e^{be}e^{cf}-e^{bc}e^{ef}\right)h_{be}h_{cf}\\
&=\bar{a}^{6}+\epsilon \bar{a}^{4}h +\frac{\epsilon^{2}\bar{a}^{2}}{2}\left(h^{2}-h^{ab}h_{ab}\right).
\end{split}
\end{equation}
We can calculate $q^{\pm\frac{1}{2}}$ using a Taylor expansion to second order in perturbations. We list the results below:
\begin{subequations}
\begin{align}
\sqrt{q}&=\bar{a}^{3}+\frac{\epsilon \bar{a}h}{2}+\frac{\epsilon^{2}}{8\bar{a}}\left(h^{2}-2h^{ab}h_{ab}\right)\\
\frac{1}{\sqrt{q}}&=\frac{1}{\bar{a}^{3}}- \frac{\epsilon h}{2\bar{a}^{5}}+\frac{\epsilon^{2}}{8\bar{a}^{7}}\left(h^{2}+2h^{ab}h_{ab}\right).
\end{align}
\end{subequations}
We will now calculate the curvature terms. It is natural to start with the Christoffel symbols
\begin{equation}
\begin{split}
\Gamma^{a}_{bc}&=\epsilon\frac{q^{ad}}{2}\left(h_{bd,c}+h_{cd,b}-h_{bc,d}\right)\\
&=\epsilon\left[\frac{e^{ad}}{2\bar{a}^{2}}\left(h_{bd,c}+h_{cd,b}-h_{bc,d}\right)\right]-\epsilon^{2}\left[\frac{h^{ad}}{2\bar{a}^{4}}\left(h_{bd,c}+h_{cd,b}-h_{bc,d}\right)\right]
\end{split}
\end{equation} 
where every partial derivative is spatial. The three Ricci scalar is:

\begin{equation}
\begin{split}
R^{(3)}=&\frac{\epsilon}{\bar{a}^{4}}\left(\partial_{a}\partial_{b}h^{ab}-\partial^{2}h\right)+\frac{\epsilon^{2}}{\bar{a}^{6}}\left[h^{ab}\partial_{a}\partial_{b}h+h^{ab}\partial^{2}h_{ab}-2h^{ai}\partial_{b}\partial_{i}h_{a}{}^{b}-\left(\partial_{a}h^{ab}\right)\left(\partial_{c}h_{b}{}^{c}\right)\right]\\
&+\frac{\epsilon^{2}}{\bar{a}^{6}}\left[\left(\partial_{a}h^{ab}\right)\left(\partial_{b}h\right) -\frac{1}{4}\left(\partial_{a}h\right)\left(\partial^{a}h\right)+\frac{3}{4}\left(\partial_{c}h^{ab}\right)\left(\partial^{c}h_{ab}\right)-\frac{1}{2}\left(\partial_{c}h^{ab}\partial_{a}h_{b}{}^{c}\right)\right].
\end{split}
\end{equation}

Now we will calculate the momentum terms. We start with the specification for $\pi^{ab}$:
\begin{equation}
\pi^{ab}=\frac{\bar{p}}{6\bar{a}}e^{ab}+\epsilon p^{ab}.
\end{equation}
We will calculate $\pi$ in detail; we start with the definition:
\begin{equation}
\begin{split}
\pi&=q_{ab}\pi^{ab}\\
&=\left(\bar{a}^{2}e_{ab}+\epsilon h_{ab}\right)\left(\frac{\bar{p}}{6\bar{a}}e^{ab}+p^{ab}\right)\\
&=3\bar{a}^{2}\left(\frac{\bar{p}}{6\bar{a}}\right)+\epsilon\left[\bar{a}^{2}p+\left(\frac{\bar{p}}{6\bar{a}}\right)h\right]+\epsilon^{2}h_{ab}p^{ab}.
\end{split}
\end{equation}
A similar calculation for $\pi_{ab}$ reveals:
\begin{equation}
\begin{split}
\pi_{ab}&=\left(a^{2}\right)^{2}\left(\frac{\bar{p}}{6\bar{a}}\right)e_{ab}+\epsilon\left[2a^{2}\left(\frac{\bar{p}}{6\bar{a}}\right)h_{ab}+\left(a^{2}\right)^{2}p_{ab}\right]+\epsilon^{2}\left[2\bar{a}^{2}p^{d}{}_{(a}h_{b)d}+\left(\frac{\bar{p}}{6\bar{a}}\right)h_{a}{}^{d}h_{bd}\right].
\end{split}
\end{equation}

We substitute these results in the expression for the Hamiltonian density, expand to second order in perturbations and simplify where possible using integration by parts. The curvature and kinetic terms from \eqref{eq:grham} are respectively:
\begin{subequations}
\begin{align}
\begin{split}
-\sqrt{q}R^{(3)}=&\frac{h^{ab}}{2\bar{a}^{3}}\left(\partial_{b}\partial^{c}h_{ac}-\frac{\partial^{2}h_{ab}}{2}\right)-\frac{h}{2\bar{a}^{3}}\left(\partial_{a}\partial_{b}h^{ab}-\frac{\partial^{2}h}{2}\right)
\end{split}\\
\frac{\pi_{ab}\pi^{ab}}{\sqrt{q}} - \frac{\pi^{2}}{2\sqrt{q}}=&\frac{1}{\bar{a}}\left(\frac{\bar{p}}{6\bar{a}}\right)\left(p^{ab}h_{ab}-\frac{hp}{2}\right)+\bar{a}\left(p^{ab}p_{ab}-\frac{p^{2}}{2}\right)+\frac{1}{8\abar^3}\left(\frac{\bar{p}}{6\bar{a}}\right)^{2}\left(5h_{ab}h^{ab}-\frac{3h^{2}}{2}\right)
\end{align}
\end{subequations}

 \section{Derivation of graviton equation}\label{gravitonA}
 
The graviton equations are those for the phase space variables $h_I(k,t)$ and $p_I(k,t)$ for  $I=3,4$ derived from the Hamiltonian (\ref{g-H}):
\begin{subequations}
\begin{align}
\dot{h}_{I}&=\frac{2}{\abar}\left[\left(\frac{\bar{p}}{6\abar}\right)h_{I}+\frac{p_{I}}{\abar^2}\right], \label{eq:h3d}\\
\dot{p}_{I}&=\frac{\bar{a}^{3}}{2}\left[-\frac{k^{2}}{\bar{a}^{2}}-\frac{5}{2\bar{a}^{2}}\left(\frac{\bar{p}}{6\bar{a}}\right)^{2}+V(\bar{\phi})-\frac{\bar{p}_{\phi}^{2}}{2\bar{a}^{6}}\right]h_{I}-\frac{2}{\bar{a}}\left(\frac{\bar{p}}{6\bar{a}}\right)p_{I}\label{eq:p3d}.
\end{align}
\end{subequations}
The first of these gives   
\begin{equation}
\label{eq:p3}
p_I=\abar^2 \left[\frac{\abar}{2}\dot{h}_I -\left(\frac{\pbar}{6\abar}\right)h_I \right],
\end{equation} 
and 
\begin{equation}
\ddot{h}_{I}=\frac{1}{3\bar{a}^{2}}\left(\dot{\bar{p}}-2\bar{p}\bar{H}\right)h_{I}+ \left(\frac{\pbar}{3\abar^2}\right)\dot{h}_{I}-\frac{6\bar{H}}{\abar^3}p_I+\frac{2}{\abar^3} \dot{p}_I,
\end{equation}
where $\bar{H} \equiv \dot{\abar}/\abar = -\pbar/12\abar^2$ from the equations for the background. 
Substituting for $p_I$ and $\dot{p}_I$  into the last equation gives  
\begin{equation}
\ddot{h}_{I}=\left[\frac{1}{3\bar{a}^{2}}\left(\dot{\bar{p}}+\pbar\bar{H}\right)-
\frac{k^{2}}{\bar{a}^{2}}+\frac{3}{2\bar{a}^{2}}\left(\frac{\bar{p}}{6\bar{a}}\right)^{2}+V(\bar{\phi})-\frac{\bar{p}_{\phi}^{2}}{2\bar{a}^{6}}\right]h_{I}-3\bar{H}\dot{h}_{I}.
\end{equation}
Finally using the background equation (\ref{eq:zeroeqns}) for $\dot{\pbar}$ gives
\begin{equation}
\ddot{h}_{I}  + 3\bar{H}\dot{h}_{I} + \left( \frac{k}{\abar} \right)^2h_{I} =0.
\end{equation}

 \section{Diffeomorphism constraint is first class}\label{diffA}

  To show that the diffeomorphism constraint  $\tilde{C}_\parallel$ (\ref{sc-diff}) is first class we must show that 
\begin{equation}
\label{eq:fcc}
\frac{d\tilde{C}_\parallel}{dt}=\{\tilde{C}_\parallel,H^{S\phi}\}+\frac{\partial \tilde{C}_\parallel}{\partial t} =0.
\end{equation}  
 The first term is  
\be
\label{eq:peom}
\{\tilde{C}_\parallel,H^{S\phi}\}&=&-2\bar{a}^{2}\left(\dot{p}_{1}+\sqrt{2}\dot{p}_{2}\right)+\left(\frac{\bar{p}}{6\bar{a}}\right)\left(\dot{h}_{1}-2\sqrt{2}\dot{h}_{2}\right)+\sqrt{3}\bar{p}_{\phi}\dot{\tphi}\nn\\
&=&-\frac{\bar{p}}{3}\left(p_{1}+\sqrt{2}p_{2}\right)-\left[\frac{1}{4a}\left(\frac{\bar{p}}{6\bar{a}}\right)^{2}+\frac{1}{\bar{a}}\left(\frac{\bar{p}_{\phi}}{2\bar{a}}\right)^{2}-\frac{\bar{a}V(\bar{\phi})}{2}\right]\left(h_{1}-2\sqrt{2}h_{2}\right)\nn\\
&&+\sqrt{3}\bar{a}^{3}V'(\bar{\phi})\tphi,
\ee
and  the second term is 
\begin{equation}
\label{eq:beom}
\frac{\partial \tilde{C}_\parallel}{\partial t}=-4\bar{a}\dot{\bar{a}}\left(p_{1}+\sqrt{2}p_{2}\right)+\left(\frac{\dot{\bar{p}}}{6\bar{a}}-\frac{\dot{\bar{a}}\bar{p}}{6\bar{a}^{2}}\right)\left(h_{1}-2\sqrt{2}h_{2}\right)+\sqrt{3}\dot{\bar{p}}_{\phi}\tphi.
\end{equation}
Substituting into this the equations for the background (\ref{eq:zeroeqns})  and collecting terms gives 
\begin{equation}
\frac{d\tilde{C}_\parallel}{dt}=0.
\end{equation}
Similar steps  show that  the same results holds for the transverse components of the linearized diffeomorphism constraint.

\bibliography{dustC}

\end{document}